# Bitcoin Mining Decentralization via Cost Analysis

Jonathan Harvey-Buschel[*], Can Kisagun[+]

[*]MIT EECS, [+]MIT Sloan

**Abstract**

Bitcoin mining presents a significant economic incentive for efficient hashing and broadcast of data, both parameters stemming from the Proofs-of-Work used to advance the network. This incentive has led to the development of Bitcoin-specific application-specific integrated circuits (ASICs) and centralized mining pools, undermining the decentralized motivations behind Bitcoin's design. In addition, the imminent block reward halving threatens the profitability of mining at any scale. Some work has been done in formal models for miner profitability, but existing models do not account for conditions such as the pricing of off-peak power and diverse investment strategies regarding sunken costs. There is also a lack of formal study of how the profit model changes as mining scales from the individual to the industrial level. Given the lack of analysis of these conditions, there are alternative models for profitable or net-zero mining that operate at smaller, and therefore more desirable, scale.

**1. Introduction**

The intent of our work is to improve the health of the Bitcoin network. Some metrics for network health include number of miners on the network, number of operating pools, percentage of hashrate per pool, and total hashrate. Ideally, the network hashrate would comprise of many low-hashrate actors participating in a diverse set of pools with low individual percentages of the network hashrate. As a lower bound for decentralization, no one actor, miner or pool, should control 25% or more of the hashrate [1]. This distribution is dependent on the accessibility and profitability of mining. Bitcoin mining is founded on submitting Proof-of-Work with a difficulty that increases to keep the time to a solution roughly constant. Generating this proof is computationally intensive, and can be modelled as a conversion of electricity to heat [2]. In the Bitcoin case, the economic efficiency of proof generation is determined by electricity costs, hardware efficiency, and operational costs such as cooling. Framed in this way, it is clear that Bitcoin mining is most efficient where it is easiest to dissipate power or cheapest to consume it. These two properties are in conflict; heat dissipation is difficult with high densities of generation, but electricity pricing favors large loads. A third key factor is processing efficiency; as the hardware involved has moved past general-purpose processors to modern ASICs, the cost for miners to stay competitive has increased rapidly. This is caused by increasing engineering costs for new ASICs, as well as growing incentives for miners to keep ASICs they design to themselves as a competitive advantage. The development of this hardware and its impact on the network is clear when observing network hashrate over time:

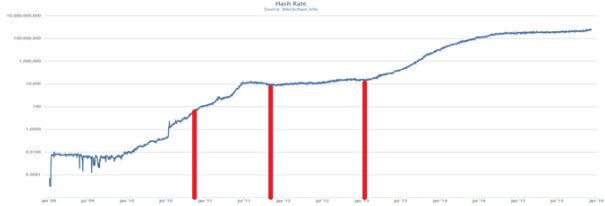

*Figure 1. Network Hashrate over Time* [3]

Figure 1 depicts the network hashrate over time. The x axis denotes time, and the y axis denotes network hashrate on a logarithmic scale. Each vertical line in this figure marks the release to the public of a significantly new development in Bitcoin mining hardware. The first line marks cgminer, which allowed enthusiasts to mine Bitcoin using their graphics cards (GPUs) [4]. Prior to this development, mining was limited to then central processing unit (CPU), a processor type common to all computers. GPUs are a popular computer component, but develop more rapidly than CPUs and use much more power. GPUs perform orders of magnitude better than CPUs at mining because they have many parallel cores designed for repeated mathematical operations on changing data, making CPU mining unprofitable. The second mark signals the development of Bitcoin-specific field-programmable gate arrays (FPGAs), which are chips that can be programmed to model any processor design. [5]. FPGAs and ASICs, which represent the last mark on Figure 1, provided further performance by implementing many parallel cores dedicated to only computing SHA256 hashes, the main operation involved in Bitcoin mining. FPGAs and ASICs are both expensive dedicated hardware that most participants in the network would not own outside of the purpose of Bitcoin mining. The resulting hashrate increase made GPU mining irrelevant, again raising the threshold for profitable participation in mining. The hashrate contributed by new hardware made GPU miners a much smaller percentage of the network, and consequently greatly reduced the reward for these miners. The corresponding change in number of miners is shown in Figure 2.

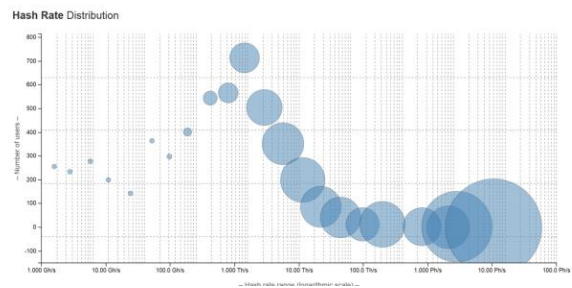

*Figure 2. Number of Miners over Time* [6]

In Figure 2, the x axis represents the network hashrate on a logarithmic scale, and the size of the circle matches the value on the x axis. The y axis represents the number of miners, which serves as a measure of centralization. A lower number indicates more centralization and vice versa. It is clear from Figure 2 that mining decentralization has already peaked, and at the moment mining is highly centralized with a persistent trend towards further centralization. The peak in this figure corresponds to the time period just before the release of FPGAs to the public.

In addition to hashrate and available hardware, bitcoin price fluctuations play a significant role in determining how decentralized the network hashrate will be. Miner rewards are denoted in bitcoin, but miner costs are paid in a fiat currency such as the U.S. Dollar or Chinese Yuan. Therefore, the conversion rate between bitcoin and fiat sets a bound for miner profitability. With an inflated bitcoin price, miners need not be very efficient, as their reward is also inflated. This also accelerates a difficulty increase, as more equipment comes online in response to larger profits. In the other case, a depressed price forces inefficient miners to take a loss, and only those who have enough fiat savings to weather a depressed conversion rate can sustain operations during these times. It is common amongst larger miners to mine at a loss during these times, and hold mined bitcoin until prices rise again.

This effect, combined with the recurring sunken cost of new ASICs, keeps miner incentives tied to the Bitcoin price over long periods of time. Miners are subject to a recurring sunken cost from investing in ASIC developments focused on efficiency, and therefore further competition. There are not many significant ASIC developments left, as the processes used have caught up to the limits of fabrication for any processor type. With these trends in mind, we seek to identify a way to increase mining decentralization through operating cost reduction, which can lower the threshold for new participation in Bitcoin mining.

## 2. Prior Literature

Prior work has been done addressing the fundamental principles of Proof-of-Work, centralizing forces in mining, and profit models for Bitcoin miners. Poelstra and Swanson both identify electricity costs & physics limitations of hardware as key considerations with respect to the evolution of the network [7] [2]. Their theses are consistent with assumptions stated earlier regarding the limits of dissipating heat in concentrated areas, and also the conversion of electricity to heat by hardware.

With regard to centralization risks, the *Mind the Gap* presentation by Carlsten, Kalodner, and Narayanan addresses the increasing time-dependence of miner profits, and therefore behavior. [9] As miner revenues over time shift from the block reward towards transaction fees, mining will only be profitable when the sum of transaction fees for unconfirmed transactions on the network exceeds the miner threshold of cost to mine. This oscillating hashrate introduces major security risks for the network, as an attacker could control a percentage of the network hashrate much smaller than 51% if attacking during an unprofitable time on the network. The Selfish Mining attacks outlined by Ittay Eyal and Emin Gun Sirer also become more attractive under these conditions [1]. One way to address this risk is with miner profitability dependent on cycles external to the Bitcoin network, such as electricity prices.

Prior work also includes economic modeling of Bitcoin mining, as seen in the *Minting Money with Megawatts* presentation. [10] This group developed a function to define miner profit, seen here:

$$\pi(X) = \frac{X}{h_0 + X} \times B \times (S + F) - X \times C - \frac{1}{T} \times \left(\frac{X}{z} + NRE\right)$$

*Figure 4. Existing Miner Profit Function* [10]

The first term is a product of percentage of network hashrate and reward denominated in U.S. dollars. The second and third terms adjust for operating and sunken costs involved in mining. The function from Figure 4 is represented graphically in Figure 5:

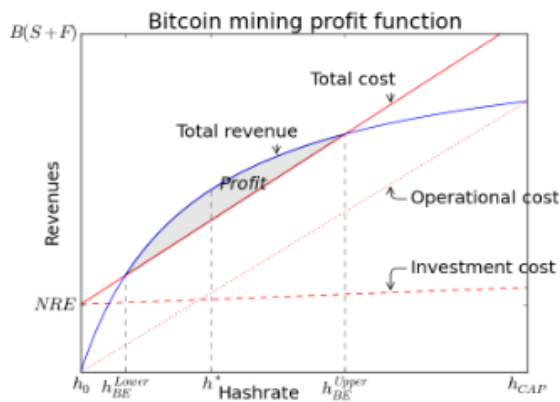

*Figure 5. Existing Miner Profit Model* [10]

Figure 5 measures hashrate on the x axis, with revenues on the y axis. The lower value of non-recurring engineering (NRE) refers to non-recurring cost of engineering, and the upper value of conversion rate multiplied by block reward and fee payments (B(S+F)) refers to the reward per block in USD. This model assumes a non-recurring cost of engineering, as well as an operational cost that scales linearly with hashrate. Based on the data we collected, we have revised this to reflect sunken and operational costs that are not linear with respect to hashrate. Our model is further modified by not accounting for miners designing their own hardware, but instead buying from existing ASIC vendors and making their first investment in mining. We also consider operational cost to scale non-linearly, due to the shift from lack of overhead at the smallest scale to datacenter design, construction, and administration at the largest scale [11] [12] [13].

## 3. Definitions

Bitcoin mining can be modeled as a conversion of electricity to heat. The cost of mining is dependent on energy efficiency, electricity pricing, cost to exhaust heat, and sunken costs such as new infrastructure and labor. Bitcoin mining can be expanded in a more distributed and cost-effective manner through widespread access to variable electricity pricing schemes and hardware setups small enough to not require additional infrastructure.

Through analysis of historical trends of the mining industry and collection of data from current miners, we have identified three categories into which all present-day mining fits. The main distinction between categories is electricity consumption. We define a Hobbyist miner as characterized by not mining with any access to electricity outside of their living space. This constrains them to running hardware in their home, and thus caps the maximum amount of electricity they can consume. Homes are only wired for consumption on the order of 10s of kilowatts (kW), and are limited to residential electricity prices. A Hobbyist would therefore run from 1 to 7 units with a 10 kW limit. The only notable infrastructure cost is that of a smart meter, the utility of which is expanded upon is Sections 4 and 6. A 2 thousand USD smart meter leaves sunken costs in the range of 3 to 9 thousand USD.

The next category of miner is the Semi-Professional; they are using a dedicated space for mining with specialized high-voltage equipment. This miner has a power envelope of 50 to 250 kW. This distinction is made because at this size, the Semi-Professional will not be directly connected to the grid or have to deal with the highest voltages used in modern grids. The level of infrastructure and design is still relatively low. A Semi-Professional miner could run 30 to 150 units, and other investments such as networking gear and racks lead to a sunken cost of 40 to 200 thousand USD, not including the cost of space used to host the equipment.

The final classification of miner is the Professional. These miners have an electricity demand of 1 megawatt (MW) or more, and therefore would run 500 or more units. These miners are designing spaces for maximum density and cooling, and are able to connect directly to low-cost electricity sources such as hydroelectric dams [7] [14]. These operations also include labor as an operating cost, and easily cost upwards of 1 million USD not including space. Even within this category there are different styles of miner based on sunken cost dedicated to optimization [11].

Currently, the supermajority of the current network hashrate is controlled by Professional miners [16]. This is a product of ASIC manufacturers becoming exclusive miners, an activity excluded from this model. Given the current state of mining, the objective is to reduce the return on investment (ROI) period for smaller actors.

For the purpose of these calculations, the Bitmain Antminer S7 is used as the model ASIC [15]. This ASIC was chosen because it is one of few publicly available, and therefore accessible to miners across segments. A unit is defined as including an ASIC and power supply. The cost per unit is about 1 thousand USD, but varies depending on size of miner [15] [11] [13] [12]. With Semi-Professional miners, ASIC orders are discounted by about 0.5% per unit, and Professional-sized orders of 300+ units are discounted by 1.5%+ [15] [13]. The cost per MW of infrastructure ranges from 50 to 250 thousand USD, depending on the level of optimization [11]. Infrastructure for miners includes networking, power distribution, cooling, and physical layout. Infrastructure does not include costs to step down high-voltage power sources, which is essential if establishing a new multi-MW facility. That cost is on the order of 100s of thousands of USD per MW [12]. The cost to purchase facilities is also not included, but inherently non-existent for Hobbyist miners.

Given these figures for the scaling of sunken costs, the percentage of cost dedicated to non-ASIC equipment grows from 20% for the Hobbyist to >50% for the Professional. This increasing sunken cost is yet another argument for Hobbyist mining; the Hobbyist model by definition is more efficient with respect to hashrate per dollar invested.

**4. Electricity Markets**

Electricity markets are highly complex and this complexity results in major inefficiencies that Bitcoin miners can leverage, including Multi-party Supply Chains, Market Features, and Demand-based Pricing Schemes.

There are a large number of actors involved in electricity generation, transmission, and consumption. As is seen in Figure 7, supply side power generators supply electricity to an Independent System Operator (ISO), the market making and price setting entity in the electricity market. Utilities and Energy Service Companies (ESCOs) purchase electricity capacity from ISOs and transmit it to their customers.

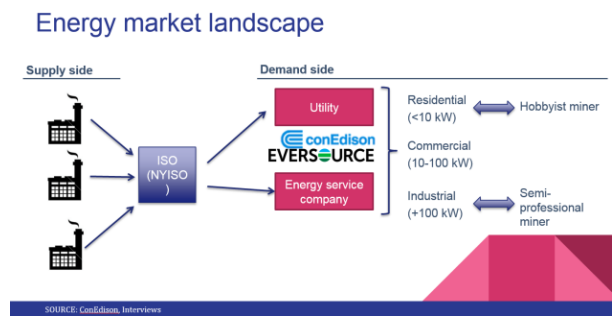

*Figure 7. Electricity market landscape* [17] [18]

There are two electricity markets: the Day-Ahead Market, which accounts for 95% of electricity purchased, and the Real-Time market [17]. Prices are set on an hourly basis a day in advance for the Day-Ahead Market and in real-time on 5-minute intervals for the Real-Time Market. Our analysis is focused on Day-Ahead Pricing for wholesale electricity market prices.

Customers are divided into three segments based on their electricity consumption: residential customers, commercial customers and industrial customers. Hobbyist miners are residential electricity customers and Semi-Professional miners are industrial electricity customers. Actors with significant electricity demands can surpass dealing with utilities or ESCOs, form Load Serving Entities (LSEs), and buy directly from wholesale markets. In the case of New York, an LSE can only serve customers within the jurisdiction of the NYISO. An example of is the agreement between financial institution Fidelity and the New England ISO to power Fidelity's data centers [18].This direct purchase model offers a saving opportunity for semi-professional miners in certain regions if they consolidate their electricity demand and buy electricity directly from ISOs and dis-intermediate Utilities and ESCOs.

The second feature of electricity markets that Bitcoin miners can take advantage of is the spread between electricity production and consumption prices. As a commodity, the electricity price is subject to demand and supply. ISOs run a uniform clearing auction to set the prices in the wholesale market. In a Day-Ahead market, suppliers submit hourly production bids at certain prices (e.g. 100MW/h for $2,000). Demand side stakeholders, such as utilities and ESCOs, submit their hourly demand forecasts and upper bounds for prices. ISOs set hourly prices for the next day based on previous demand and supply trends. ISOs select the lowest cost suppliers and continue with higher cost suppliers until production meets demand. The price at which production meets demand is the hourly price in the Day-Ahead Market [17].

Since electricity cannot be stored in large quantities, the on-demand generation must supply system demand to avoid outages. This results in discrepancies between forecasted demand from the previous day and the actual demand. There also exists a real-time market price for electricity, decided in 5 minute intervals, for players wishing to purchase electricity at the last moment. Real-Time Markets work with the same economic principles as the Day-Ahead Market. However, prices in the Real-Time Market are much more volatile. Due to the clearing auction nature of price setting, the cost of generating energy increases in high demand periods. When there is extra demand (peak demand), suppliers operate more expensive electricity resources such as coal plants or gas combustion turbines. As seen in Figure 8, the real cost of serving the customer is highly variable and is much higher during the peak demand hours. The x axis displays hours of the day, and the y axis displays the cost per kilowatt-hour (kWh) during that hour.

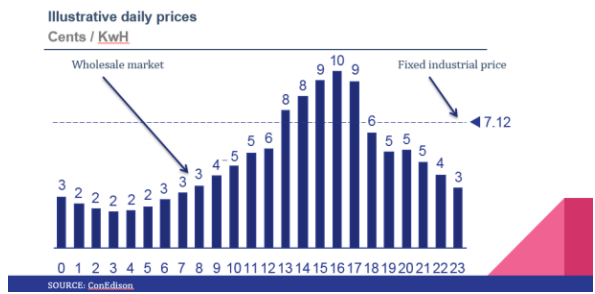

*Figure 8. Electricity Costs* [18]

Most existing retail customers are on fixed price schemes, while larger customers use Time of Use (TOU) tariffs. Fixed price schemes involve an insurance premium for utilities and ESCOs as these players have to hedge against the price and demand fluctuations [18]. This insurance premium is passed on to the consumers. Figure 9 shows fixed and TOU price schemes offered by ConEdison for the analysis period for residential customers in New York. The x axis displays hours of the day, and the y axis displays the cost per kilowatt-hour (kWh) during that hour.

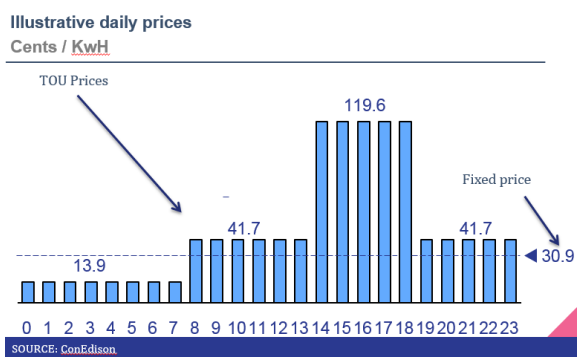

*Figure 9. Daily Electricity Prices* [19]

Recently utilities have made an effort to accelerate the deployment of Smart Meters, which allow electricity users to access TOU pricing schemes. As of 2009, private and public utilities in Oregon, Idaho, California, Arizona, Texas, Oklahoma, Florida, Alabama, Georgia, South Carolina, Wisconsin, Indiana, Michigan, Ohio, Pennsylvania, Massachusetts and Vermont had +50% smart meter deployment plans. In California alone, utilities SCE, PG&E and SDG&E deployed 11.8mn smart meters. Massachusetts Green Community Act mandates a state-wide pilot and a state-wide deployment of potentially 2.6mn smart meters [18]. The potential to access significantly cheaper electricity at certain times of the day or with certain power offers miners new means to reduce the payback period for hardware.

## 5. Methodology

The methodology used involved revised revenue and cost models, and applying them with a backwards-looking analysis using network data such as hashrate, price, transaction fees, and network difficulty. The analysis was performed on data from August 1st to September 20th, 2015, and December 1st of 2015 to January 20th, 2015. The first period is characterized by a slowly-changing difficulty, decreasing network hashrate, and a price oscillating around 250 USD. The second period is notably different, with the network hashrate growing by about 66%, difficulty growing by 50%, and a price moving around 400 USD. The case of price data from the first period during the network state of the second period is also considered to simulate a significant miner reward drop. The analysis uses a simple model for miner setup based on current publicly available hardware, as explained earlier. Our revised revenue function follows:

$$P_{day} = \left(\frac{X}{h_0 + X}\right) * B(S + F) * \frac{(h_0 + X) * (10^9 * 86400)}{D * 2^{32}}$$

X – Personal Hashrate (GH/s)

$H_0$ – Network Hashrate (GH/s)

B – Bitcoin Price (USD)

S – Block Reward

F – Fee Volume

D – Network Difficulty

The $\frac{X}{h_0 + X}$ term represents the miner's percentage of the total network hashrate, and the product with the $B(S + F)$ returns revenue in USD. The $\frac{(h_0+X)*(10^9*86400)}{D*2^{32}}$ term is used to adjust for the network difficulty and number of blocks generated per day. This value changes both with the difficultly readjustment every two weeks and the daily network hashrate fluctuation. Our revised cost function follows:

$$Cost_{hour} = X * time * P_{electricity}$$

This definition of operating cost fluctuates based on the price of electricity, which varies greatly based on location, time of day, and level of access to electricity. The primary cost reduction for Semi-Professional miners will be derived from this fluctuation, and other defining features of electricity markets.

Electricity price data is limited to New York, specifically ConEdison and the NYISO. Other state and country electricity price data is considered in Section 7.

## 6. Analysis of Results

Our analysis is structured to test the impact of the two saving opportunities we identified for miners in the section above. Figure 10 below shows average cost of mining for different miner segments and mining methods.

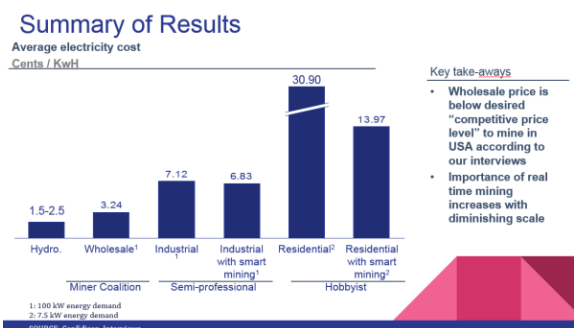

*Figure 10. Summary of Results* [20]

The values for each column represent average electricity cost in cents per kWh. The first saving opportunity is the ability to purchase electricity directly from ISOs. This brings down the average cost of electricity to 3.24 cents per kWh, which is below the "acceptable mining cost threshold" of 4 cents per kWh based on our interviews [12]. However, total demand required to buy directly from wholesale markets vary between 1MW and 3MW based on different ISOs. Only Professional miners will be able to negotiate access to these prices, and miners experience difficultly obtaining access due to a lack of established trust between miners and electricity companies. Not only are miners a new type of high-demand electricity client, but there have been multiple incidents where irresponsible miners have significantly damaged the grids they were connected to because of poor management [14]. Within the first analysis period, halving electricity costs only resulted in a 15% increase in profit margin. For the second period, the increase was also 15%. Using price data from the first period on the network state of the second, this profit increase doubles to 31%. Alternatively, with a 12.5 bitcoin block reward at the same price as the first period, profits increase by 42% instead of 14%.

The second saving opportunity is the use of data on real-time electricity prices to calculate expected profit on an hourly basis, and set the power state of mining hardware based on the expected profit. Real-time mining algorithms can calculate hourly expected revenue based on the hashrate of the individual miner, hashrate of the network, current network difficulty, current bitcoin prices, and hourly electricity costs. Currently, real time pricing (RTP) is possible either through RTP schemes offered by certain utilities such as Georgia Power [18]. These programs are based on Day-Ahead wholesale market prices and are set on an hourly basis. ConEdison, the major utility in our analysis region of New York, does not offer real-time pricing but instead offers TOU schemes. Thus the smart-mining cost analysis for industrial and residential customers are driven from TOU pricing scheme for respective segment.

Lower average electricity costs with smart mining are derived from non-continuous mining, due to negative revenues at certain times of day. Based on the ConEdison TOU scheme, the impact of real-time mining for Semi-Professional miners is about 1%. This effect is not only small, but also infeasible given the rapid powering on and off of mining equipment. The larger the mine, the more time needed to safely power on and off the equipment. This 1% increase also holds for the second period. For Hobbyist miners, the impact of real-time mining is significant. Real-time mining enables Hobbyists to access 50% cheaper average electricity prices and improve expected profits by a multiple of 7 during the first period. Units would be powered on for only 59% of the day compared to the standard 100%. 100% duty cycle represents always-on mining. As represented in Figure 11, Hobbyists using smart mining can increase their expected profit from $137 to $929.

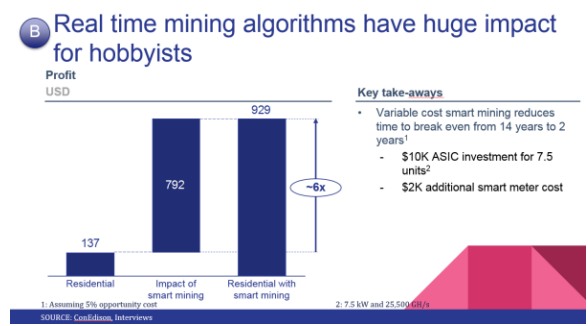

*Figure 11. Hobbyist Real-Time Results* [21]

During the second period, this multiplier decreases to 4.5. ASIC daily duty cycle increases from 59% to 61%. Most interesting is the case of the price data of the first period with the network state of the second, as real-time mining in this case switches Hobbyist mining from a highly unprofitable activity to a profitable one. This case represents a miner reward halving, without a block reward halving.

**7. Discussion**

These results prove that there is a significant opportunity for increased decentralization of mining through operating cost optimization at small scales. While Semi-Professional miners do not benefit significantly from access to wholesale electricity market prices, Hobbyist miners can reclaim a reasonable ROI by accessing variable electricity prices. The Hobbyist savings on infrastructure and operating cost make this model attractive not just from a profit-focused perspective, but also considering the metrics of network health. The Real-Time Hobbyist model is valid anywhere that a resident has access to variable electricity prices, which applies to many developed nations. EU-15 countries such as the UK and Spain have access to variable pricing, and electricity is much more expensive in Europe compared to the US [23]. Excluding Hawaii and Alaska, all regions in the US have lower residential rates than those in EU-28 countries excluding Bulgaria [24]. Thus, the Real-Time Hobbyist model is more attractive to a hobbyist miner in Europe. Also, countries that produce electricity locally, such as OPEC states, provide sizable electricity subsidies (sub-cent/kWh) to residents and small businesses, allowing for profitable decentralized fixed-price always-on mining [25].

With a current average network hashrate of more than 1 Exahashes per second (EH/s), a 5% participation would be 50 Petahashes per second (PH/s). This translates to 70 Semi-Professional miners running about 700

Terahashes per second (TH/s) each, or about 1,600 Hobbyist miners running 30 TH/s each, following our models of the maximum size of each miner type. The upfront total investment for this would be about $15 million USD. A 5% hashrate is similar in size to the current fifth-biggest pool hashrate [16].The rough consensus concerning equilibrium hashrate is that mining activity will eventually converge to only the locations where it is profitable or investments have been paid off [7] [2]. These locations drop in number as the block reward drops, which is the supermajority (99 %+) of miner profits at the moment. This profitability and consequent centralization patterns are heavily affected by existing resources such as cheap power and inefficient electricity grids. What we have proven is that, for both historically low miner profitability, turbulent network conditions, and a combination of the two, Hobbyists can significantly lower their operating costs via real-time mining and greatly lower the threshold for profitable mining.

Despite the positive takeaways from our work, there are some significant limitations to this model. The most significant is that we performed a backwards looking analysis, instead of a forwards looking analysis. This is primarily because Bitcoin as a market and network is very difficult to understand and predict the future behavior of, given how many external factors affect it. For this practical reason, we chose to use historical network data we knew was valid. A significant rise in Bitcoin prices reduces the appeal for cost savings as the threshold for profitable mining drops.

Another assumption we made was that new miners were only concerned with real-time profits, i.e. were "shorting" Bitcoin and valuing it only with respect to its current price. This is contrary to the standard Professional miner behavior, but allows for better measurement of the impact of cost saving techniques. Those willing to mine at a loss could still use these techniques, but they are less appealing if the goal is Bitcoin-denominated profits vs. fiat profits. We proved the utility of our method for the "short" case – The utility diminishes if the miner is more invested in Bitcoin and is willing to absorb some operating costs to hold their reward as Bitcoin. However, this model now gives the Hobbyist a choice; they can increase their duty cycle from the ~60% we found upwards to 100% at any given time to reflect their long-term investment in Bitcoin.

**6. Future Work**

With regards to future work, we plan to conduct interviews with ISOs to better understand the exact requirements for buying electricity directly from wholesale markets. This includes identifying the minimum required demand, and cost and processes involved in purchasing electricity directly. A case study focusing on energy-rich countries with strict capital controls such as Venezuela would assist in proving the social utility of the Real-Time Hobbyist miner model. Countries such as Ecuador, Paraguay, Mozambique,

Burma, and Vietnam are net energy exporting countries (+100 trillion BTU) with at least 5% of energy production driven from renewable sources, meriting future study regarding their energy infrastructure and the potential role PoW hosting could play [27].

Another worthwhile cost model to consider would be sunken vs. operating costs, such as investing in renewable energy. With access to variable electricity costs, a Hobbyist could dynamically and efficiently allocate grid-sourced and self-sourced electricity to essential (appliances, lights, heating) and auxiliary (Bitcoin ASICs, Tor Nodes) electrical loads. If renewable energy networks are sized for peak demand, but their production matches day/night cycles, there may be a similar opportunity to absorb excess electricity from the grid. A similar study for running nodes would also be useful, as proving a lower operating cost could have a similar effect for those considering joining the network. The health of the Bitcoin network depends on many related metrics, but through more efficient essential resource usage (electricity, storage, and bandwidth), we can further develop the decentralization of Bitcoin.